\documentstyle[12pt,epsfig]{article}
\begin{document}
\baselineskip = 0.24 in
\topmargin= -15mm
\textheight= 230mm

\begin{center}
{\Large  {\bf New spectrum and condensate in two dimensional QCD}}
\end{center}

\vspace{0.5cm}

\begin{center}
T. Fujita\footnote{e-mail: fffujita@phys.cst.nihon-u.ac.jp},
M. Hiramoto\footnote{e-mail: hiramoto@phys.cst.nihon-u.ac.jp} and
T. Homma\footnote{e-mail: hommaj@phys.cst.nihon-u.ac.jp}\\
Department of Physics, Faculty of Science and Technology \\
Nihon University, Tokyo, Japan
\end{center}

\vspace{0.5cm}

\begin{center}
\section*{\large Abstract} 
\end{center}

The boson mass and the condensate in two dimensional QCD with $SU(N_c)$ colors 
are numerically evaluated with the Bogoliubov vacuum. 
It is found that the boson mass 
is finite at the massless limit, and it is well described by the 
phenomenological expression 
$ {\cal M}_{N_c}={2\over 3}\sqrt{{N_cg^2\over{3\pi}}} $ for large $N_c$. 
Also, the condensate values agree very 
well with the prediction by the $1/N_c$ expansion. The validity 
of the naive light cone method is examined, and it turns out that the light 
cone prescription of the boson mass with the trivial vacuum 
is accidentally good for QED$_2$. 
But it is not valid for QCD$_2$. Further, at the massless fermion limit, the chiral 
symmetry is spontaneouly broken without anomaly, and the Goldstone 
theorem 
does not hold for QCD$_2$.

\vspace{0.5cm}

PACS numbers:  11.10.Kk, 11.15.-q, 03.70.+k, 11.30.-j, 11.30.Rd \par

\newpage

\section{Introduction}
The physics of the strong interaction is described by Quantum Chromodynamics (QCD). 
Yet, it is extremely difficult to solve the theory in a nonperturbative way  
and obtain any reasonable spectrum of bosons. In this respect, it is quite interesting 
to understand QCD with  one space and one time dimensions (QCD$_2$) since this 
field theory model can be solved in a nonperturbative fashion. 

The boson spectrum in QCD$_2$ has been extensively studied 
by the light cone method \cite{q01,q022,q02}. 
In particular, QCD$_2$ with the $1/N_c$ expansion proposed by 't Hooft has 
presented  interesting results on the boson mass 
spectrum \cite{q21,q210,q211,q22,q23}. The boson mass 
vanishes when the fermion mass becomes zero. However, this is not allowed since 
the massless boson cannot {\it physically} exist 
in two dimensional field theory \cite{q3,q4}. 
Unfortunately, this problem of the puzzle has never been seriously considered 
until now, apart from unrealistic physical pictures. People believe that 
the large $N_c$ limit is special because one takes $N_c$ infinity. 
But the infinity in physics means simply that the $N_c$ must be 
sufficiently large, and, in fact, as we show below, physical observables 
at $N_c=50$ are just the same as those of $N_c = \infty$. 

Further, this boson spectrum of large $N_c$ QCD$_2$ was confirmed 
by the light cone calculations 
with $SU(2)$ and $SU(3)$ colors \cite{q022}. 
Indeed, the mass of the boson in the light cone calculations 
is consistent with the 't Hooft spectrum of the boson even though the latter is 
evaluated by the $1/N_c$ approximation. However, the fact that the light cone calculation 
predicts massless bosons is rather serious since 
the light cone calculation for $SU(2)$ does not seem to make any unrealistic 
approximations, apart from the trivial vacuum. 

However, there is an interesting indication that the light cone vacuum is 
not trivial, and indeed there is a finite condensate even for the large 
$N_c$  QCD$_2$ \cite{q05,q06,q03,q04}. What does this mean ? 
This suggests that one has to consider 
the effect of the complicated vacuum structure for the boson mass as long as 
one calculates the boson mass  with Fock space expansions. On the other hand,  
the calculation for the boson spectrum by 't Hooft is based on the trivial vacuum,  
but, instead he could sum up all of the intermediate 
fluctuations of the fermion and antifermion pairs. This should be 
equivalent to considering the true vacuum structure in the Fock space basis. 
That is, the same spectrum of bosons must be obtained both by 
the Fock space expansion with the true vacuum and by the sum of 
all the Feynman diagrams with the trivial vacuum if they are 
treated properly. 

For this argument, people may claim that QED$_2$ is exactly described by the 
naive light cone calculation with the trivial vacuum, and therefore, QCD$_2$ 
may well be treated just in the same way as the QED$_2$ case. However, 
one may well have some uneasy feeling for the fact that the naive 
light cone calculation cannot reproduce the condensate value of QED$_2$. 

In this paper, we show that the light cone calculation based 
on the Fock space expansion with the trivial 
vacuum is not valid for QCD$_2$. One has to consider properly the effect 
of the complicated vacuum structure. Here, we present the calculation with 
the Bogoliubov vacuum in the rest frame, 
and show that the present calculation reproduces the right condensate values. 
Indeed, we can compare the present results with the condensate value  
as predicted by the $1/N_c$ expansion\cite{q05,q06,q03,q04}, 
$$ C_{N_c}=-{N_c\over{\sqrt{12}}} \sqrt{N_cg^2\over{2\pi}} . \eqno{(1.1)} $$ 
The present calculation of the condensate value for the $SU(2)$ color 
is $C_2=-0.495$  ${g\over{\sqrt{\pi}}} $ which should be 
compared with the $-0.577$  ${g\over{\sqrt{\pi}}} $ from the $1/N_c$ expansion, and 
$C_3=-0.995$  ${g\over{\sqrt{\pi}}} $ for the $SU(3)$ color  compared 
with $-1.06$  ${g\over{\sqrt{\pi}}} $ of the $1/N_c$ expansion. 
For the larger value of $N_c$ (up to $N_c=50$), we obtain the condensate values 
which perfectly agree with the prediction of $C_{N_c}$ in eq.(1.1). 

Further, we show that the boson masses for QCD$_2$ 
with $SU(2)$ and $SU(3)$ colors are finite 
even though the fermion mass is set to zero. 
In fact, the boson mass is found 
to be ${\cal M}_2=0.467$ ${g\over{\sqrt{\pi}}} $ for the $SU(2)$, 
and ${\cal M}_3=0.625$ ${g\over{\sqrt{\pi}}} $ 
for the $SU(3)$ color for the massless fermions. 
Further, the present calculations of the boson mass up to $N_c=50$ 
suggest that the boson mass ${\cal M}_{N_c}$ for $SU(N_c)$ can be described 
for the large $N_c$ by the following phenomenological expression at 
the massless fermion limit, 
$$  {\cal M}_{N_c}={2\over 3}\sqrt{{N_cg^2\over{3\pi}}} . \eqno{(1.2)} $$
Also, we calculate the boson mass at the large $N_c$ with the finite 
fermion mass. From the present calculations, we can express the boson mass 
in terms of the phenomenological formula for the small fermion mass $m_0$ region, 
$$  {\cal M}_{N_c} \approx \left( {2\over 3}{\sqrt{2\over 3}}
+{10\over{3}}{m_0\over{\sqrt{N_c}}} \right) 
\sqrt{{N_cg^2\over{2\pi}}}  \eqno{(1.3)} $$
where $m_0$ is measured in units of ${g\over{\sqrt{\pi}}}$. 

The above expression (eq.(1.3)) can be compared with the calculation 
by Li et al. \cite{q210} who employed the $1/N_c$ expansion of 't Hooft model 
in the rest frame \cite{q211}. 
It turns out that their calculated boson mass for their smallest fermion 
mass case is consistent with the above equation, though their calculated 
values are slightly smaller than the present results. 
   
In addition, we examine the validity of the light cone calculation 
for QED$_2$. It is shown that the boson mass for the QED$_2$ case 
happens to be not very sensitive to the condensate value, and that 
the spectrum can be reproduced by the light cone calculations 
with the trivial vacuum as well as with the condensate value only 
with positive momenta. 
Therefore, we believe that the QED$_2$ case is accidentally reproduced 
by the light cone calculation with the trivial vacuum state even though 
we do not fully understand why this accidental agreement occurs. 
On the other hand, the QCD$_2$ case is quite different. 
The boson mass calculated with the trivial 
vacuum is zero at the massless fermion limit. Further, the calculation 
in the light cone with the condensate value only with the positive momenta 
are not stable against the infra-red singularity of the light cone equations. 

The present calculations are based on the Fock space expansion, and, in this 
calculation, we only consider the fermion and anti-fermion (two fermion) 
space.  For QED$_2$, it is shown that the fermion and anti-fermion 
space reproduces the right Schwinger boson \cite{q11}. 
That is, the four fermion spaces 
do not alter the lowest boson energy in QED$_2$. However, there is 
no guarantee that there are finite effects on the lowest boson mass 
from the four fermion spaces in QCD$_2$. This point is not examined 
in this paper, and should be worked out in future. 

Here, we examine the RPA calculations and show that the boson mass 
for QED$_2$ with the RPA equations deviates from the Schwinger boson. 
That means that the agreement achieved by the Fock space expansion 
is destroyed by the RPA calculation. 
Further, we calculate the boson mass for QCD$_2$ 
with $SU(2)$ and the large $N_c$ limit. It turns out that the boson 
mass vanishes when the fermion mass is equal to the critical value 
and that it becomes imaginary when the fermion mass is smaller than 
the critical value. This is obviously unphysical at the massless fermion limit, 
and  is closely related to the fact 
that the RPA equations are not Hermitian, 
and therefore  we should examine its physical meaning in future. 

From the present calculation, we learn that the chiral symmetry in massless  QCD$_2$ 
is spontaneouly broken without the anomaly term, in contrast to the Schwinger model. 
But the boson mass is finite, and therefore 
there is no Goldstone boson in this field theory model. 
Thus, the present result confirms that 
the Goldstone theorem \cite{q1,q2} does not hold for the fermion field theory 
as proved in ref. \cite{q9}. This indicates that the anomaly term has 
little to do with the chiral symmetry breaking. This is reasonable since 
the anomaly term arises from the conflict between the gauge invariance 
and the chiral current conservation when regularizing the vacuum, 
and this is essentially a kinematical effect. 
On the other hand, the symmetry breaking is closely related to the lowering 
of the vacuum energy, and therefore it is the consequence of 
the dynamical effects in the vacuum. 

This paper is organized as follows. In the next section, we briefly 
explain the Bogoliubov vacuum in QCD$_2$, and obtain the equations for 
the condensate as well as for the boson mass. In sections 3 and 4, we examine 
the boson masses for QCD$_2$ and for QED$_2$ in the light cone calculation, 
respectively. In section 5, we examine the problem of the 't Hooft 
calculation of the boson mass in large $N_c$ limit of QCD$_2$. 
Section 6 treats the RPA calculations in QED$_2$ and QCD$_2$, and 
section 7 discusses the spontaneous chiral symmetry breaking in QCD$_2$. 
In section 8, we summarize what we have clarified in this paper.

\section{Bogoliubov transformation in QCD$_2$}
In this section, we discuss the Bogoliubov transformation in QCD$_2$. 
The Lagrangian density for QCD$_2$ with $SU(N_c)$ color is described as 
$$ {\cal L}= 
\bar{\psi}(i\gamma^{\mu}\partial_{\mu}-g\gamma^{\mu}A_{\mu} -m_0)\psi 
-\frac{1}{4}F^{a}_{\mu\nu}F^{a\mu\nu},
 \eqno{(2.1)} $$
where $F_{\mu\nu}$ is written as
$$ F_{\mu\nu} = \partial_{\mu}A_{\nu}-\partial_{\nu}A_{\mu}+ig[A_{\mu},A_{\nu}] $$
$$
A_{\mu} = A^{a}_{\mu}T^{a}, \ \ \  
T^{a} = \frac{\tau^{a}}{2} .  $$ 
$m_0$ denotes the fermion mass, and at the massless limit, the Lagrangian density 
has a chiral symmetry.

Now, we first fix the gauge by 
$$ A^a_1 =0 . \eqno{(2.2)} $$
In this case, the Hamiltonian of QCD$_2$ with $SU(N_c)$ color  can be written as 
$$ H = \sum_{n,\alpha}p_{n}
\left(a_{n,\alpha}^{\dagger}a_{n,\alpha}-b_{n,\alpha}^{\dagger}b_{n,\alpha}
\right)  
 +m_0 \sum_{n,\alpha}
\left(a_{n,\alpha}^{\dagger}b_{n,\alpha}+b_{n,\alpha}^{\dagger}a_{n,\alpha}
\right)  $$
$$ -\frac{g^2}{4N_cL}\sum_{n,\alpha,\beta}\frac{1}{p^{2}_{n}}
\left({\tilde j}_{1,n,\alpha\alpha}+{\tilde j}_{2,n,\alpha\alpha}
\right)
\left({\tilde j}_{1,-n,\beta\beta}+{\tilde j}_{2,-n,\beta\beta}
\right)
\nonumber $$
$$ +\frac{g^2}{4L}\sum_{n,\alpha,\beta}\frac{1}{p^{2}_{n}}
\left({\tilde j}_{1,n,\alpha\beta}+{\tilde j}_{2,n,\alpha\beta}
\right)
\left({\tilde j}_{1,-n,\beta\alpha}+{\tilde j}_{2,-n,\beta\alpha}
\right), \eqno{(2.3)} $$
where
$$ 
{\tilde j}_{1,n,\alpha\beta} = 
\sum_{m}a_{m,\alpha}^{\dagger}a_{m+n,\beta} \eqno{(2.4a)} $$
$$ {\tilde j}_{2,n,\alpha\beta} = 
\sum_{m}b_{m,\alpha}^{\dagger}b_{m+n,\beta} . \eqno{(2.4b)} $$
Now, we define new fermion operators by the Bogoliubov transformation, 
$$ a_{n,\alpha}=\cos\theta_{n,\alpha}c_{n,\alpha}+\sin\theta_{n,\alpha}
d_{-n,\alpha}^{\dagger} \eqno{(2.5a)} $$
$$
b_{n,\alpha}=-\sin\theta_{n,\alpha}c_{n,\alpha}+\cos\theta_{n,\alpha}
d_{-n,\alpha}^{\dagger} \eqno{(2.5b)} $$
where $ \theta_{n,\alpha}$ denotes the Bogoliubov angle. 

In this case, the Hamiltonian of QCD$_2$ can be written as 
$$ H = \sum_{n,\alpha}E_{n,\alpha}(c^\dagger_{n,\alpha}c_{n,\alpha}
+d^\dagger_{-n,\alpha}d_{-n,\alpha}) +H' \eqno{(2.6)} $$
where
$$ E^2_{n,\alpha} = \left\{p_{n}+\frac{g^2}{4N_cL}\sum_{m,\beta}
\frac{(N_c\cos 2\theta_{m,\beta}-\cos 2\theta_{m,\alpha})}{(p_m-p_n)^2}
\right\}^2 $$
$$ +\left\{m_0+ \frac{g^2}{4N_cL}\sum_{m,\beta}{
(N_c\sin 2\theta_{m,\beta}-\sin 2\theta_{m,\alpha}) 
\over{{(p_m-p_n)^2} }} \right\}^2 .
 \eqno{(2.7)} $$
$H'$ denotes the interaction Hamiltonian in terms of the new operators but 
is quite complicated, and therefore it is given in Appendix. 

The conditions that the vacuum energy is minimized give the constraint 
equations which can determine the Bogoliubov angles 
$$ \tan 2\theta_{n,\alpha} =
\frac{m_0+\frac{g^2}{4N_cL}\sum_{m,\beta}
\frac{(N_c\sin 2\theta_{m,\beta}-\sin 2\theta_{m,\alpha})}
{(p_m-p_n)^2}}
{p_{n}+\frac{g^2}{4N_cL}\sum_{m,\beta}\frac{(N_c\cos 2\theta_{m,\beta}-
\cos 2\theta_{m,\alpha})}{(p_m-p_n)^2}} . \eqno{(2.8)} $$
In this case, the condensate value $C_{N_c}$ is written as 
$$ C_{N_c} ={1\over L} \sum_{n,\alpha} \sin 2\theta_{n,\alpha} . \eqno{(2.9)} $$
Now, we can calculate the boson mass for the $SU(N_c)$ color. First,  we 
define the wave function for the color singlet boson as 
$$ |\Psi_{K}\rangle = \frac{1}{\sqrt{N_c}}\sum_{n,\alpha}f_{n}
c^\dagger_{n,\alpha}d^\dagger_{K-n,\alpha}|0\rangle .  \eqno{(2.10)} $$
In this case, the boson mass can be described as 
$$ {\cal M} = \langle \Psi_{K}|H|\Psi_{K}\rangle
 = \frac{1}{N_c}\sum_{n,\alpha}\left(E_{n,\alpha}
+E_{n-K,\alpha}\right)|f_n|^2 $$
$$
+\frac{g^2}{2N_c^2L}\sum_{l,m,\alpha}\frac{f_{l}f_{m}}{(p_l-p_m)^2}
\cos(\theta_{l,\alpha}-\theta_{m,\alpha})
\cos(\theta_{l-K,\alpha}-\theta_{m-K,\alpha}) $$
$$-\frac{g^2}{2N_cL}\sum_{l,m,\alpha,\beta}\frac{f_{l}f_{m}}{(p_l-p_m)^2}
\cos(\theta_{l,\alpha}-\theta_{m,\beta})
\cos(\theta_{l-K,\alpha}-\theta_{m-K,\beta}) $$
$$ +\frac{g^2}{2N_c^2L}
\sum_{l,m,\alpha,\beta}\frac{f_{l}f_{m}}{K^2}
\sin(\theta_{l-K,\alpha}-\theta_{l,\alpha})
\sin(\theta_{m,\beta}-\theta_{m-K,\beta}) $$
$$ -\frac{g^2}{2N_cL}
\sum_{l,m,\alpha}\frac{f_{l}f_{m}}{K^2}
\sin(\theta_{l-K,\alpha}-\theta_{l,\alpha})
\sin(\theta_{m,\alpha}-\theta_{m-K,\alpha}) . \eqno{(2.11)} $$
This equation can be easily diagonalized together with the Bogoliubov 
angles, and we obtain the boson mass. Here, we note that the treatment 
of the last two terms should be carefully estimated since the apparent 
divergence at $K=0$ is well defined and finite.

\subsection{  Condensate and boson mass in $SU(2)$ and $SU(3)$ }

Here, we present our calculated results of the condensate values 
and the boson mass in QCD$_2$ with the $SU(2)$ and $SU(3)$ colors. 
Table 1 shows the condensate and the boson mass for the two 
different vacuum states, one with the trivial vacuum and 
the other with the Bogoliubov vacuum. As can be seen, the 
condensate values for the $SU(2)$ and $SU(3)$ are already close 
to the predictions  by the $1/N_c$ expansion of eq.(1.1) \cite{q05,q06,q04}. 
The boson masses for the $SU(2)$ and $SU(3)$  are for the first time 
obtained as the finite value. Unfortunately, we cannot compare our 
results with any other predictions. But compared with the Schwinger 
boson, the boson masses for the $SU(2)$ and $SU(3)$ are in the same order 
of magnitude. 

In fig. 1, we present the fermion mass dependence of the condensate 
values for the $SU(2)$ and $SU(3)$ cases. As can be seen, the condensate 
becomes a finite value at the massless limit with the linear 
dependence on the fermion mass $m_0$. 
Also, in fig. 2, we show   
the calculated results of the boson mass as the function of the $m_0$ 
for $SU(2)$ and $SU(3)$. At the massless limit, the boson mass 
becomes a finite value, and the $m_0$ dependence is linear. 
This is exactly the same as the QED$_2$ case \cite{q11,q15}. 

The present calculations show that both of the 
values (condensate and boson mass) are  a smooth function 
of the fermion mass $m_0$. This means 
that the vacuum structure has no singularity at the massless limit. 
This must be due to the fact that the coupling constant $g$ has 
the mass dimension and therefore, physical quantities are expressed 
by the coupling constant $g$ even at the massless limit 
of the fermion. This is in contrast to the Thirring model 
where the massless limit is a singular point. 
In the Thirring model, the coupling constant has no dimension, and 
therefore, at the massless limit, physical quantities 
must be described by the cutoff $\Lambda$.

\vspace{0.2cm}
\begin{center}
Table 1 \\
\ \ QCD$_2$ with $SU(2)$ and $SU(3)$ in rest frame  \\ 
in units of \  ${g\over{\sqrt{\pi}}}$ with $m_0=0$ \\
\ \ \\
\begin{tabular}{|c||c|c|c|}
\hline
\ $SU(2)$ & Trivial  & Bogoliubov  & $1/N_c$  \\
\hline
\hline
Condensate & 0 & $-0.495$ & $-0.577$  \\
\hline
Boson Mass & $-\infty$    & 0.467  & 0  \\
\hline
\end{tabular} 

\vspace{0.5cm}

\begin{tabular}{|c||c|c|c|}
\hline
\ $SU(3)$ & Trivial  & Bogoliubov  & $1/N_c$  \\
\hline
\hline
Condensate & 0 & $-0.995$ & $-1.06$  \\
\hline
Boson Mass & $-\infty$   & 0.625  & 0  \\
\hline
\end{tabular} 
\end{center}

In Table 1, we show the condensate values and the boson mass 
of QCD$_2$ in the rest frame.  
Here, the minus infinity of the boson mass in the trivial vacuum 
is due to the mass singularity $ \ln (m_0) $ as explained in ref. \cite{q11}

\vspace{0.5cm}

\subsection{  Condensate and boson mass in $SU(N_c)$ }

Here, we  carry out the calculations of the condensate 
and the boson mass for the large $N_c$ values of $SU(N_c)$  
up to $N_c=50$. In fig. 3, we show the calculated condensate values 
(denoted by crosses) as the function 
of the $N_c$ together with the prediction of the $1/N_c$ expansion 
as given in eq.(1.1). As can be seen, the calculated condensate values 
agree very well 
with the prediction of the $1/N_c$ expansion if the $N_c$ is larger than 10. 
Further, the calculated  boson masses (denoted by crosses) 
are shown in fig. 4 as the function of $N_c$. 
It is found that they can be described by the following formula [eq.(1.2)] 
for the large $N_c$ values, 
$$  {\cal M}_{N_c}={2\over 3}\sqrt{{N_cg^2\over{3\pi}}} . \eqno{(1.2)} $$ 
Indeed, the calculated boson masses for $N_c $ larger than 
$N_c =10 $ perfectly agree with the predicted value of eq.(1.2). 

The present calculations show that the second excited state for $SU(N_c)$ 
colors is higher than the twice of the boson mass at the massless fermions. 
Therefore, there is only one bound state in QCD$_2$ with the $SU(N_c)$. 
This indicates that eq.(1.2) must be the full boson spectrum 
for QCD$_2$ with massless fermions. 

Now, we present the calculations of the boson mass for the finite fermion mass 
$m_0$ cases. Here, we limit ourselves to the $m_0$ (in units of ${g\over{\sqrt{\pi}}}$) 
which is smaller than unity. 
In fig. 5, we show the calcualted  values of the boson mass 
as the function of  ${m_0\over{\sqrt{N_c}}}$ for several cases of the fermion 
mass $ m_0 $ and the color $N_c$. The present calculation is 
carried out up to the $N_c=50$ case which is sufficiently large enough 
for the large $N_c$ limit of the 't Hooft model. 
The solid line in fig. 5 is obtained as the following phenomenological 
formula of the fit to the numerical data 
$$  {\cal M}_{N_c} \approx \left( {2\over 3}{\sqrt{2\over 3}}
+{10\over{3}}{m_0\over{\sqrt{N_c}}} \right) 
\sqrt{{N_cg^2\over{2\pi}}} . \eqno{(1.3)} $$
Now, we want to compare the present results with the old calculations 
by Li et al. who obtained the boson mass by solving the 't Hooft equations 
for QCD$_2$ with the large $N_c$ limit in the rest frame. 
Li et al. obtained the boson mass for their smallest fermion mass 
of $ m_0=0.18 \sqrt{{N_c\over{2}}} $
$$  {\cal M}_{\infty} = 0.88 \sqrt{{N_cg^2\over{2\pi}}} . \eqno{(2.12)} $$
There are also a few more points of their calculations with larger fermion 
mass cases. In fig. 5, we plot the boson masses calculated by Li et al. by 
the white circles which should be compared with the solid line. 
As can be seen, the boson mass obtained by 
Li et al. is close to the present calculation. It should be noted that 
their calculations were carried out with rather small number 
of the basis functions in the numerical evaluation, and therefore, the 
accuracy of their calculations may not be very high, in particular, for 
the small fermion mass regions.  

Unfortunately, however, Li et al. made a wrong conclusion 
on the massless fermion limit 
since their calculated point of $ m_0=0.18 \sqrt{{N_c\over{2}}} $  
was the smallest fermion mass. Obviously, this value of the fermion mass 
was by far too large to draw any conclusions on the massless fermion limit.

\vspace{1cm}

\section{    QCD$_2$ in light cone }

Here$B!$(Bwe evaluate the boson mass in the light cone. For this, we follow 
the prescription in terms of the infinite momentum frame \cite{q12,q15} 
since this has a good connection to the rest frame calculation. 
In this frame, we can calculate the boson mass with and without 
the condensate in the light cone. But in evaluating the condensate, 
we only consider the positive momenta. 
The equation for the boson mass square for the $SU(2)$ case becomes 
$$ {\cal M}^2 = m^2_{0}\int dxf(x)^2\left(\frac{1}{x}+\frac{1}{1-x}\right) $$
$$ +\frac{3g^2}{16\pi}\int dxdy
\frac{f(x)^2}{(x-y)^2}
\left(\cos 2\theta_{y,1}+\cos 2\theta_{1-y,1}  
 +\cos 2\theta_{y,2}+\cos 2\theta_{1-y,2}\right) $$
$$ -\frac{g^2}{4\pi}\int dxdy\frac{f(x)f(y)}{(x-y)^2}
\Big[\frac{1}{2}\cos(\theta_{x,1}-\theta_{y,1})
\cos(\theta_{x-1,1}-\theta_{y-1,1}) $$
$$ +\frac{1}{2}\cos(\theta_{x,2}-\theta_{y,2})
\cos(\theta_{x-1,2}-\theta_{y-1,2}) 
 +\cos(\theta_{x,1}-\theta_{y,2})\cos(\theta_{x-1,1}-\theta_{y-1,2}) $$
$$ +\cos(\theta_{x,2}-\theta_{y,1})\cos(\theta_{x-1,2}-\theta_{y-1,1})
\Big] $$
$$ -\frac{g^2}{8\pi}\int dxdyf(x)f(y)
\Big[\sin(\theta_{x,1}-\theta_{x-1,1})
\sin(\theta_{y-1,1}-\theta_{y,1}) $$
$$ +\sin(\theta_{x,2}-\theta_{x-1,2})
\sin(\theta_{y-1,2}-\theta_{y,2})  $$
$$ -\sin(\theta_{x,1}-\theta_{x-1,1})
\sin(\theta_{y-1,2}-\theta_{y,2}) 
 -\sin(\theta_{x,2}-\theta_{x-1,2})
\sin(\theta_{y-1,1}-\theta_{y,1}) \Big]  \eqno{(3.1)} $$ 

Here, all of the momenta are positive. 
This can be easily evaluated, and we obtain the condensate 
values and the boson mass as given in Table 2. 
We note here that both of the values become smaller 
as the function of the fermion mass, and finally they 
vanish to zero. This is exactly what is observed in the light 
cone calculations. Since the light cone calculations cannot reproduce 
the condensate values which are finite as predicted in ref. \cite{q03}, 
the light cone calculations must have some problems. 
In Table 2, we also show the calculations of the infinite momentum frame 
with the positive momenta only. However, the numerical calculations are 
not stable against the infra-red singularity of the light cone. 
At the present stage, we do not know how to evaluate them properly, and 
we do not fully understand what is wrong with the light cone. 

\vspace{0.5cm}

\begin{center}
Table 2 \\
\ \ QCD$_2$ with $SU(2)$ in infinite momentum frame \\ 
in units of \  ${g\over{\sqrt{\pi}}}$  with $m_0=0$ \\
\ \ \\
\begin{tabular}{|c||c|c|c|}
\hline
\ $SU(2)$ & Trivial  & Bogoliubov ($p>0$)  & $1/N_c$  \\
\hline
\hline
Condensate & 0 & $**$  & $-0.577$  \\
\hline
Boson Mass & $0$   & $**$  & 0  \\
\hline
\end{tabular} 
\end{center}

\vspace{1cm}

\section{  Examinations of  QED$_2$ in light cone }

In this section, we examine the validity of the light cone calculation 
for the boson mass in  QED$_2$. It is well known that the light cone calculation 
for the boson mass gives an exact result  of the Schwinger model \cite{q02,q15}. 
However, it is also confirmed that the vacuum of QED$_2$ should possess 
a finite condensate value even in the light cone vacuum. This indicates that 
the agreement of the light cone calculation for the boson mass with the trivial vacuum 
may well be accidental. 

Here, we examine the light cone calculation for the boson mass by taking into account 
the effect of the condensate of the vacuum in QED$_2$. Since all of the equations 
for QED$_2$ are just the same as QCD$_2$ case, we present here only the calculated 
results for the boson mass in the light cone. In this calculation, the condensate 
of the vacuum is estimated only by the positive momenta of the vacuum state. 

Before presenting the light cone results, we first show the calculation 
of the rest frame with the trivial vacuum and the Bogoliubov vacuum states. 
In Table 3a, we show the condensate value and the boson mass with the 
two vacuum states. As can be seen, the trivial vacuum can reproduce 
neither the condensate value nor the boson mass. On the other hand, the Bogoliubov 
vacuum can reproduce both the right condensate and the right boson mass \cite{q11}. 

Now, we go to the infinite momentum frame. 
In Table 3b, we show the condensate value and the boson mass. 
It is surprising to see that the right boson mass is reproduced by the calculation 
with the trivial vacuum even though the condensate is not reproduced. 
However, the infinite momentum frame calculations with the Bogoliubov vacuum 
with the positive momenta only again have the numerical instabilty 
against the infra-red singularity of the light cone. 
Nevertheless, the right boson mass is reproduced at the small fermion mass. 

This is somewhat puzzling, and it looks that the boson mass 
in QED$_2$ is insensitive to the condensate.

In Table 3, we show the condensate values and the boson mass 
of QED$_2$ in the rest frame 
and in the infinite momentum frame calculations.

\vspace{2cm}
\begin{center}
Table 3a \\
\ \ QED$_2$ in rest frame  \\ 
in units of \  ${g\over{\sqrt{\pi}}}$ with $m_0=0$ \\
\ \ \\
\begin{tabular}{|c||c|c|}
\hline
\ QED$_2$ & Trivial Vac & Bogoliubov Vac   \\
\hline
\hline
Condensate & 0 & $-0.283$  \\
\hline
Boson Mass & $-\infty$    & 1.0   \\
\hline
\end{tabular} 
\end{center}

\vspace{0.5cm}

\begin{center}
Table 3b \\
\ \ QED$_2$  in infinite momentum frame \\ 
in units of \  ${g\over{\sqrt{\pi}}}$  with $m_0=0$ \\
\ \ \\
\begin{tabular}{|c||c|c|}
\hline
\ QED$_2$ & Trivial Vac & Bogoliubov Vac   \\
\hline
\hline
Condensate & 0 & $ $**$ $   \\
\hline
Boson Mass & 1.0  & 1.0  \\
\hline
\end{tabular} 
\end{center}

\vspace{1cm}

\section{  Examination of  't Hooft model }

Here, we discuss the boson mass of QCD$_2$ with $SU(N_c)$ color 
in the large $N_c$ limit. 
This model is solved by 't Hooft who sums up all of the Feynman diagrams 
in the $1/N_c$ expansion and obtains the equations for the boson mass. 
In principle, the 't Hooft equations must be exact up to the order of $1/N_c$. 
Therefore, one does not have to consider the effect of the vacuum since 
the 't Hooft equations take into account all of the fluctuations of the intermediate 
fermion and antifermion pairs. 
Therefore, it is expected that the right boson mass can be obtained 
from the equations at the order of $1/N_c$. 

The present calculations of the boson mass with the $SU(N_c)$ colors 
show that the boson mass can be well described by 
$ {\cal M}_{N_c}={2\over 3}\sqrt{{N_cg^2\over{3\pi}}} $  as the function of $N_c$ 
for the large values of $N_c$. 
In the 't Hooft model, the boson mass should be proportional to 
$\sqrt{N_cg^2\over{2\pi}}$, and therefore, the present expression of the boson mass 
is consistent with the 't Hooft evaluation 
as far as the expansion parameter is concerned. 
Therefore,  the boson mass calculation by the planar diagram evaluations 
of  't Hooft must be reasonable.   

Therefore, the boson mass prediction of 't Hooft should be reexamined 
from the point of view of  the light cone procedure.  It seems that the 
 't Hooft equations in the light cone have lost one important 
information which is expressed in terms of the $\theta_p$ variables 
both in the paper by Bars and Green \cite{q211} and also in the present paper. 
Since the variables $\theta_p$ are closely  related to the condensate 
values, the equations without the $\theta_p$ variables should correspond 
to the trivial vacuum in our point of view. Therefore, if one can recover 
this constraint in the 't Hooft equations in the light cone, 
then one may obtain the right boson mass from the  't Hooft model. 

\vspace{0.5cm}

\section{ RPA calculations in  QED$_2$ and  QCD$_2$}

Up to this point, we have presented the calculated results of the Fock space 
expansion with the Bogoliubov vacuum state for  QED$_2$ and  QCD$_2$. 
The lowest  boson mass which is calculated by the Fock space expansion 
must be exact for the fermion and anti-fermion states 
if the vacuum is exact. From the present result for the condensate values 
of  QED$_2$ and  QCD$_2$, it indicates that the Bogoliubov vacuum state 
should be very good or may well be exact. 

On the other hand, there are  boson mass calculations by employing 
the Random Phase Approximation (RPA) method, and  
some people believe that the RPA calculation should be better than 
the Fock space expansion. 

Therefore, in this section, we present our calculated results of the RPA equations 
for  QED$_2$ and QCD$_2$ since there are no careful calculations in the 
very small fermion mass regions. First, we show that the RPA 
calculation for  QED$_2$ with the Bogoliubov vacuum state predicts the boson 
mass which is smaller than the Schwinger boson at the massless fermion limit. 
This means that the agreement achieved by the Fock space expansion is destroyed 
by the RPA calculation since it gives a fictitious attraction. 

Further, the RPA calculation for  QCD$_2$ with the Bogoliubov vacuum state 
produces an imaginary boson mass at the massless fermion limit. 
This is quite interesting, and it strongly suggests that the RPA equation 
cannot be reliable for fully relativistic cases 
since the eigenvalue equation of the RPA is not Hermitian,  
which is, in fact, a well known fact. 

Here, we briefly discuss the results of the RPA calculations, but  
the detailed discussion of the basic physical reason of the RPA problems 
will be given elsewhere. 

The RPA equations are based on the expectation that the backward moving 
effects of the fermion and anti-fermion may be included 
if one considers the following operator which contains 
the $d_{-m}c_m$ term in addition to the 
fermion and  anti-fermion creation term, 
$$ Q^\dagger = \sum_{n}(X_nc_n^\dagger d_{-n}^\dagger +Y_n d_{-n}c_n) . 
\eqno{(6.1)} $$
The RPA equtaions can be obtained by the following double commutations,
$$ \langle 0|[\delta Q,[H,Q^\dagger]]|0 \rangle = 
\omega \langle 0|[\delta Q,Q^\dagger]|0 \rangle  \eqno{(6.2)} $$
where $ \delta Q $ denotes 
$ \delta Q = d_{-n}c_n$ and $ c_n^\dagger d_{-n}^\dagger  $. 

Here, the vacuum $ |0 \rangle $ is assumed to satisfy the following condition, 
$$ Q |0 \rangle = 0 .  \eqno{(6.3)} $$
However, if the vacuum is constructed properly in the field theory model, 
it is impossible to find a vacuum that satisfies the condition of eq.(6.3). 
This fact leads to the RPA equations which are not Hermitian.

For  QED$_2$, the RPA equations for $X_n$ and $Y_n$ become 
$$ {\cal M} X_n = 2E_{n}X_n
-\frac{g^2}{L}\sum_{m}X_{m}
\frac{\cos^2(\theta_{n}-\theta_{m})}{(p_n-p_m)^2} $$
$$ -\lim_{\varepsilon\to 0}\frac{g^2}{L}\sum_{m}X_{m}
\frac{\sin(\theta_{n-\varepsilon}-\theta_{n})
      \sin(\theta_{m}-\theta_{m-\varepsilon})}{\varepsilon^2} $$
$$
-\frac{g^2}{L}\sum_{m}Y_{m}
\frac{\sin^2(\theta_{n}-\theta_{m})}{(p_n-p_m)^2} 
 -\lim_{\varepsilon\to 0}\frac{g^2}{L}\sum_{m}Y_{m}
\frac{\sin(\theta_{n-\varepsilon}-\theta_{n})
      \sin(\theta_{m}-\theta_{m-\varepsilon})}{\varepsilon^2}  $$
$$ -{\cal M} Y_n = 2E_{n}Y_n
-\frac{g^2}{L}\sum_{m}Y_{m}
\frac{\cos^2(\theta_{n}-\theta_{m})}{(p_n-p_m)^2} $$
$$ -\lim_{\varepsilon\to 0}\frac{g^2}{L}\sum_{m}Y_{m}
\frac{\sin(\theta_{n-\varepsilon}-\theta_{n})
      \sin(\theta_{m}-\theta_{m-\varepsilon})}{\varepsilon^2} $$
$$ -\frac{g^2}{L}\sum_{m}X_{m}
\frac{\sin^2(\theta_{n}-\theta_{m})}{(p_n-p_m)^2} 
 -\lim_{\varepsilon\to 0}\frac{g^2}{L}\sum_{m}X_{m}
\frac{\sin(\theta_{n-\varepsilon}-\theta_{n})
      \sin(\theta_{m}-\theta_{m-\varepsilon})}{\varepsilon^2} \eqno{(6.4)} $$
For QCD$_2$, one can easily derive the RPA equations, and 
at the large $N_c$ limit, they agree with the RPA equations 
which are obtained by Li et.al \cite{q210,q03}. 

It is important to note that the RPA equations 
are not Hermitian, and therefore there is no guarantee 
that the energy eigenvaules are real. In fact, as we  see below, 
the boson mass for QCD$_2$ becomes imaginary at the very small fermion mass. 

In Table 4, we show the calculated values of the boson mass 
by the RPA equations for  QED$_2$ and  QCD$_2$ with the Bogoliubov vacuum state. 
It should be noted that the boson mass for $m_0=0$ case with the Fock space 
in the large $N_c$ limit is obtained from the 't Hooft equation. This equation is 
exactly the same as eq.(2.11) if we take the large $N_c$ limit. 
We note here that the boson mass (0.543  $ \sqrt{N_cg^2\over{2\pi}}$) 
at the large $N_c$ limit with the Fock space expansion just 
agrees with the value of eq.(1.2). 

\vspace{0.5cm}

\begin{center}
{Table 4} \\
\ \ \\
The masses for QED$_2$ and  QCD$_2$ with $SU(2)$ 
are measured by $ {g\over{\sqrt{\pi}}}$. \\
The masses for large $N_c$ QCD$_2$ are measured 
by $ \sqrt{N_cg^2\over{2\pi}}$. 

\ \ \\
\begin{tabular}{|c||c|c||c|c||c|c|}
\hline
\ & \multicolumn{2}{c||}{  QED$_2$} &
\multicolumn{2}{c||}{ QCD$_2$ SU(2)}  &
\multicolumn{2}{c|}{ Large $N_c$ QCD$_2$ }   \\ 
\cline{2-7}
\ & $m_0=0$ &  $m_0 =0.1 $ &  $m_0=0$ &  $m_0 =0.1 $ 
&  $m_0=0$ &  $m_0 =0.1 $ \\
\hline
\hline
\  & \  & \  & \  & \  & \  & \   \\
Fock Space & $1.000 $ &  $1.180$ &  0.467 &  $0.709$ 
&  0.543 &  $0.783$  \\
\  & \  & \  & \  & \ & \  & \   \\
\hline
\  & \  & \  & \  & \ & \  & \   \\
RPA & $ 0.989 $ 
&   $1.172$ 
&  $0.104 i $ & $ 0.576$ &  $0.120 i $ & $ 0.614$  \\
\  & \  & \  & \  & \ & \  & \   \\

\hline
\end{tabular} \\
\end{center}

\vspace{0.5cm}

The behavior of the boson mass of the RPA calculation for  QCD$_2$ is not normal, 
contrary to the expectation. First, it is not linear as the function of $m_0$, but 
nonlinear in the small mass region. Further, the boson mass square becomes zero 
when the $m_0$ becomes a critical value, and it becomes negative 
when the $m_0$ is smaller than the critical value. In this case, 
the boson mass is imaginary, and thus this is physically not acceptable.  
This catastrophe is found to occur for the $SU(2)$ as well as 
for the large $N_c$ limit, as shown in Table 4. 

At this point, we should comment on the belief that the RPA calculation 
should produce the masssless boson at the massless fermion limit in QCD$_2$. 
However, if there were  physically a massless boson in two dimensions, 
this would be quite serious since a physical massless boson cannot propagate 
in two dimensions since it has an infra-red singularity in its propagator. 
But there is no way to remedy 
this infra-red catastrophe, and that is related to the theorem of  
Mermin, Wagner and Coleman \cite{q3,q4}. 
There are some arguments that the large $N_c$ limit is special because 
one takes the $N_c$ infinity. However, "infinity" in physics means simply 
that the $N_c$ must be sufficiently large, and in fact, as shown above, 
physical observables at $N_c=50$ are just the same as those of 
$N_c = \infty$. 
Therefore, it is rigorous that there should not exist any physical massless boson 
in two dimensions, even though one can write down the free massless boson Lagrangian 
density and study its mathematical structure. 
Thus, if one finds a massless boson constructed from the fermion and antifermion 
in two dimensions, then there must be something wrong in the calculations, 
and this is exactly what we see in the RPA calculations in  QCD$_2$. 

In this respect, the boson mass calculated only by the Fock space expansion 
with the Bogolibov vacuum can be reasonable from this point of view 
since there are some serious problems in the light cone as well as 
in the RPA calculations at the massless fermion limit. 

\vspace{1cm}

\section{  Spontaneous chiral symmetry breaking in QCD$_2$}

The Lagrangian density of QCD$_2$ has a chiral symmetry 
when the fermion mass $m_0$ is set to zero. In this case, 
there should be no condensate for the vacuum state 
if the symmetry is preserved in the vacuum state. However, as we saw 
above, the physical vacuum state in QCD$_2$ has a finite condensate value, 
and thus the chiral symmetry is broken. In contrast to the Schwinger model, 
there is no anomaly in QCD$_2$, and therefore the chiral current is 
conserved. Thus, this symmetry breaking is spontaneous. 

However, there appears no massless boson.  Even though 
no appearance of the Goldstone boson is very reasonable 
in two dimension, this means that the Goldstone theorem does not 
hold for the fermion field theory. This is just what is proved in 
ref. \cite{q10}, and the present calculations confirm the claim 
of ref. \cite{q10}. 

However, the physics of the symmetry breaking is still complicated, and 
we do not fully understand the underlying mechanism in depth. 
But it seems that the chiral anomaly does not play any important 
role in the symmetry breaking business though it has been believed 
that the Schwinger model breaks the chiral symmetry due to the anomaly. 

However, the massless limit in QED$_2$ is not singular \cite{q11}. The condensate 
value and the boson mass are smooth as the function of the fermion mass $m_0$. 
This means that the vacuum structure is smoothly connected from the massive case 
to the massless one.  

This is just in contrast to the Thirring model \cite{q5,q9,q10} where 
the massless limit is a singular point. The structure of the vacuum is 
completely different from the massive case to the massless one 
in the Thirring model. 
Further, the condensate value and the boson 
mass in the Thirring model are not smooth function of the fermion mass $m_0$. 
For the massive Thirring model, there is no condensate, and the boson mass 
is proportional to  the fermion mass $m_0$ \cite{q10,q12,q13,q15}. 
Indeed, in the massive 
Thirring model, the induced mass term arising from the Bogoliubov transformation 
is completely absorbed into the mass renormalization term, and the vacuum 
stays as it is before the Bogoliubov transformation. 
But, for the massless Thirring model, the condensate is finite, 
and the condensate value and 
the boson mass are both proportional to the cutoff $\Lambda$ by which 
all of the physical observables are measured. 

On the other hand, QED$_2$ and QCD$_2$ are very different in that 
the coupling constant of the models have the mass scale dimensions, and 
all of the physical quantities are described by the coupling constant $g$ 
even at the massless limit. 
The super-renormalizability for QED$_2$ and QCD$_2$ must be quite important 
in this respect, while the Thirring model has no dimensional quantity, and this makes 
the vacuum structure very complicated when the fermion mass is zero. 

In Table 5, we summarize the physical quantities of the chiral symmetry breaking 
for QED$_2$, QCD$_2$ and Thirring models. All the condensates and the masses 
are measured in units of ${g\over{\sqrt{\pi}}}$ for QED$_2$ and QCD$_2$. 
The $\Lambda$ and $g_0$ 
in the Thirring model denote the cutoff parameter and the coupling constant, 
respectively. Also, the value of $ \alpha (g_0) $ can be obtained by 
solving the equation for bosons in the Thirring model \cite{q9,q10}.

For QED$_2$, there is an anomaly, and therefore, the chiral current 
is not conserved while, for QCD$_2$ and the Thirring model, the chiral 
current is conserved. From Table 5, one sees that the symmetry breaking 
mechanism is just the same for QED$_2$ and QCD$_2$. However, the Thirring 
model has a singularity at the massless fermion limit, and this gives 
rise to somewhat different behaviors from the gauge theory.

\vspace{0.1cm}
\begin{center}
{Table 5} \\
\ \ \\
\begin{tabular}{|c||c|c||c|c||c|}
\hline
\ & \multicolumn{2}{c||}{  Condensate} &
\multicolumn{2}{c||}{ Boson Mass} &
 Anomaly  \\ 
\cline{2-5}
\ & $m_0=0$ &  $m_0 \not= 0$ &  $m_0=0$ &  $m_0 \not= 0$ & \  \\
\hline
\hline
\  & \  & \  & \  & \  & \    \\
QED$_2$ & $-0.283$ &  $-0.283+ O(m_0)$ &  1 &  $1+O(m_0)$ & yes  \\
\  & \  & \  & \  & \  & \    \\
\hline
\  & \  & \  & \  & \  & \    \\
QCD$_2$ & $-{N_c\over{\sqrt{12}}} \sqrt{N_c\over{2}}$ 
&   $-{N_c\over{\sqrt{12}}} \sqrt{N_c\over{2}}+O(m_0)$ & 
 ${2\over 3}\sqrt{{N_c\over{3}}}$ & $ \left( {2\over 3}{\sqrt{2\over 3}}
+{10\over{3}}{m_0\over{\sqrt{N_c}}} \right) 
\sqrt{{N_c\over{2}}}$ & no  \\
\  & \  & \  & \  & \  & \    \\
\hline
\  & \  & \  & \  & \  & \    \\
Thirring & ${\Lambda\over{g_0\sinh \left({\pi\over g_0}\right)}} $ 
& 0 &  ${\alpha (g_0)\Lambda\over{\sinh \left({\pi\over g_0}\right)}} $ 
 & $\alpha (g_0) m_0 $  & no  \\
\  & \  & \  & \  & \  & \    \\

\hline
\end{tabular} \\
\end{center}

\vspace{1cm}

\vspace{1cm}

\section{  Conclusions}

We have presented a novel calculation of the condensate and the boson mass 
for QCD$_2$ with the $SU(N_c)$ colors for the massless and massive fermions. 
The calculated condensate values $C_2$, $C_3$ and up to $N_c=50$ 
are consistent with the prediction 
$ C_{N_c}=-{N_c\over{\sqrt{12}}} \sqrt{N_cg^2\over{2\pi}}$ 
which is obtained by the $1/N_c$ expansion. In particular, the condensate 
values for the $N_c$ larger than 10 agree perfectly with the prediction. 

The boson mass for QCD$_2$ is finite, and increases as the function 
of the color $N_c$. In fact, 
the boson mass for the large $N_c$ color is obtained and expressed 
phenomenologically as 
$ {\cal M}_{N_c}={2\over 3}\sqrt{{N_cg^2\over{3\pi}}} $, and it is finite 
for the finite values of the $N_c$. This result agrees with 
the calculation of 't Hooft equations with the Fock space expansion 
in the rest frame. However, 
this contradicts the prediction of 't Hooft calculation in the light cone. 
The reason behind the disagreement may arize from the light cone 
method which is employed by 't Hooft. For this, however, we do not fully 
understand the basic reason of physics, and further studies  
are definitely needed.

\vspace{1cm}

We would like to thank  F. Lenz for helpful comments. The present  
calculations are performed with Personal Computers 
with Pentium 4 (2.8 GHz) which enables us to make complicated 
calculations which would have been very hard with 
supercomputers of five years ago.

\newpage 
\section*{ Appendix}

\begin{eqnarray}
H' &=& H_{C}+H_{A}+H_{R}+H^{(4)}+H^{(22)}.
\end{eqnarray}
\begin{eqnarray}
H_{C} &=& \frac{g^2}{4L}\sum_{n,m,l,\alpha,\beta}\frac{1}{p_n^2}
\Big[
\frac{1}{N_c}\cos(\theta_{m,\alpha}-\theta_{m+n,\alpha})
\cos(\theta_{l,\beta}-\theta_{l-n,\beta})\nonumber\\ & &
(c^{\dagger}_{m,\alpha}c_{m+n,\alpha}d^{\dagger}_{-l+n,\beta}d_{-l,\beta}
+c^{\dagger}_{l,\beta}c_{l-n,\beta}d^{\dagger}_{-m-n,\alpha}d_{-m,\alpha})
\nonumber\\
& &
-\cos(\theta_{m,\alpha}-\theta_{m+n,\beta})
\cos(\theta_{l,\beta}-\theta_{l-n,\alpha})\nonumber\\ & &
(c^{\dagger}_{m,\alpha}c_{m+n,\beta}d^{\dagger}_{-l+n,\alpha}d_{-l,\beta}
+c^{\dagger}_{l,\beta}c_{l-n,\alpha}d^{\dagger}_{-m-n,\beta}d_{-m,\alpha})\Big]
\end{eqnarray}
\begin{eqnarray}
H_{A} &=& \frac{g^2}{4L}\sum_{n,m,l,\alpha,\beta}\frac{1}{p_n^2}
\Big[
\frac{1}{N_c}\sin(\theta_{m+n,\alpha}-\theta_{m,\alpha})
\sin(\theta_{l-n,\beta}-\theta_{l,\beta})\nonumber\\ & &
(c^{\dagger}_{m,\alpha}c_{l-n,\beta}d^{\dagger}_{-m-n,\alpha}d_{-l,\beta}
+c^{\dagger}_{l,\beta}c_{m+n,\alpha}d^{\dagger}_{-l+n,\beta}d_{-m,\alpha})
\nonumber\\
& &
-\sin(\theta_{m+n,\beta}-\theta_{m,\alpha})
\sin(\theta_{l-n,\alpha}-\theta_{l,\beta})\nonumber\\ & &
(c^{\dagger}_{m,\alpha}c_{l-n,\alpha}d^{\dagger}_{-m-n,\beta}d_{-l,\beta}
+c^{\dagger}_{l,\beta}c_{m+n,\beta}d^{\dagger}_{-l+n,\alpha}d_{-m,\alpha})\Big]
\end{eqnarray}
\begin{eqnarray}
H_{R} &=& \frac{g^2}{4L}\sum_{n,m,l,\alpha,\beta}\frac{1}{p_n^2}
\Big[
\frac{1}{N_c}\cos(\theta_{m,\alpha}-\theta_{m+n,\alpha})
\cos(\theta_{l,\beta}-\theta_{l-n,\beta})\nonumber\\ & &
(c^{\dagger}_{m,\alpha}c^{\dagger}_{l,\beta}c_{m+n,\alpha}c_{l-n,\beta}
+d^{\dagger}_{-m-n,\alpha}d^{\dagger}_{-l+n,\beta}d_{-m,\alpha}d_{-l,\beta})
\nonumber\\
& &
-\cos(\theta_{m,\alpha}-\theta_{m+n,\beta})
\cos(\theta_{l,\beta}-\theta_{l-n,\alpha})\nonumber\\ & &
(c^{\dagger}_{m,\alpha}c^{\dagger}_{l,\beta}c_{m+n,\beta}c_{l-n,\alpha}
+d^{\dagger}_{-m-n,\beta}d^{\dagger}_{-l+n,\alpha}d_{-m,\alpha}d_{-l,\beta})
\Big]
\end{eqnarray}
\begin{eqnarray}
H^{(4)} &=& \frac{g^2}{4L}\sum_{n,m,l,\alpha,\beta}\frac{1}{p_n^2}
\Big[
\frac{1}{N_c}\sin(\theta_{m+n,\alpha}-\theta_{m,\alpha})
\sin(\theta_{l-n,\beta}-\theta_{l,\beta})\nonumber\\ & &
(c^{\dagger}_{m,\alpha}c^{\dagger}_{l,\beta}d^{\dagger}_{-m-n,\alpha}d^{\dagger}_{-l+n,
\beta}
+c_{m+n,\alpha}c_{l-n,\beta}d_{-m,\alpha}d_{-l,\beta})\nonumber\\
& &
-\sin(\theta_{m+n,\beta}-\theta_{m,\alpha})
\sin(\theta_{l-n,\alpha}-\theta_{l,\beta})\nonumber\\ & &
(c^{\dagger}_{m,\alpha}c^{\dagger}_{l,\beta}d^{\dagger}_{-m-n,\beta}d^{\dagger}_{-l+n,
\alpha}
+c_{m+n,\beta}c_{l-n,\alpha}d_{-m,\alpha}d_{-l,\beta})\Big]
\end{eqnarray}
\begin{eqnarray}
H^{(22)} &=& \frac{g^2}{4L}\sum_{n,m,l,\alpha,\beta}\frac{1}{p_n^2}
\Big[\frac{1}{N_c}\Big\{
\cos(\theta_{m,\alpha}-\theta_{m+n,\alpha})
\sin(\theta_{l-n,\beta}-\theta_{l,\beta})\nonumber\\ & &
(c^{\dagger}_{m,\alpha}c^{\dagger}_{l,\beta}c_{m+n,\alpha}d^{\dagger}_{-l+n,\beta}
-c^{\dagger}_{l,\beta}d^{\dagger}_{-m-n,\alpha}d^{\dagger}_{-l+n,\beta}d_{-m,\alpha}
\nonumber\\ & &
-c^{\dagger}_{m,\alpha}c_{m+n,\alpha}c_{l-n,\beta}d_{-l,\beta}
+c_{l-n,\beta}d^{\dagger}_{-m-n,\alpha}d_{-m,\alpha}d_{-l,\beta})
\nonumber\\
& &
+\sin(\theta_{m+n,\alpha}-\theta_{m,\alpha})
\cos(\theta_{l,\beta}-\theta_{l-n,\beta})\nonumber\\ & &
(-c^{\dagger}_{m,\alpha}c^{\dagger}_{l,\beta}c_{l-n,\beta}d^{\dagger}_{-m-n,\alpha}
+c^{\dagger}_{m,\alpha}d^{\dagger}_{-m-n,\alpha}d^{\dagger}_{-l+n,\beta}d_{-l,\beta}
\nonumber\\
& &
+c^{\dagger}_{l,\beta}c_{m+n,\alpha}c_{l-n,\beta}d_{-m,\alpha}
-c_{m+n,\alpha}d^{\dagger}_{-l+n,\beta}d_{-m,\alpha}d_{-l,\beta})\Big\}
\nonumber\\
& &
-\Big\{
\cos(\theta_{m,\alpha}-\theta_{m+n,\beta})
\sin(\theta_{l-n,\alpha}-\theta_{l,\beta})\nonumber\\ & &
(c^{\dagger}_{m,\alpha}c^{\dagger}_{l,\beta}c_{m+n,\beta}d^{\dagger}_{-l+n,\alpha}
-c^{\dagger}_{l,\beta}d^{\dagger}_{-m-n,\beta}d^{\dagger}_{-l+n,\alpha}d_{-m,\alpha}
\nonumber\\
& &
-c^{\dagger}_{m,\alpha}c_{m+n,\beta}c_{l-n,\alpha}d_{-l,\beta}
+c_{l-n,\alpha}d^{\dagger}_{-m-n,\beta}d_{-m,\alpha}d_{-l,\beta})\nonumber\\ & &
+\sin(\theta_{m+n,\beta}-\theta_{m,\alpha})
\cos(\theta_{l,\beta}-\theta_{l-n,\alpha})\nonumber\\
& &
(-c^{\dagger}_{m,\alpha}c^{\dagger}_{l,\beta}c_{l-n,\alpha}d^{\dagger}_{-m-n,\beta}
+c^{\dagger}_{m,\alpha}d^{\dagger}_{-m-n,\beta}d^{\dagger}_{-l+n,\alpha}d_{-l,\beta}
\nonumber\\
& &
+c^{\dagger}_{l,\beta}c_{m+n,\beta}c_{l-n,\alpha}d_{-m,\alpha}
-c_{m+n,\beta}d^{\dagger}_{-l+n,\alpha}d_{-m,\alpha}d_{-l,\beta})
\Big\}\Big]
\end{eqnarray}

\newpage
\includegraphics[width=.75\linewidth]{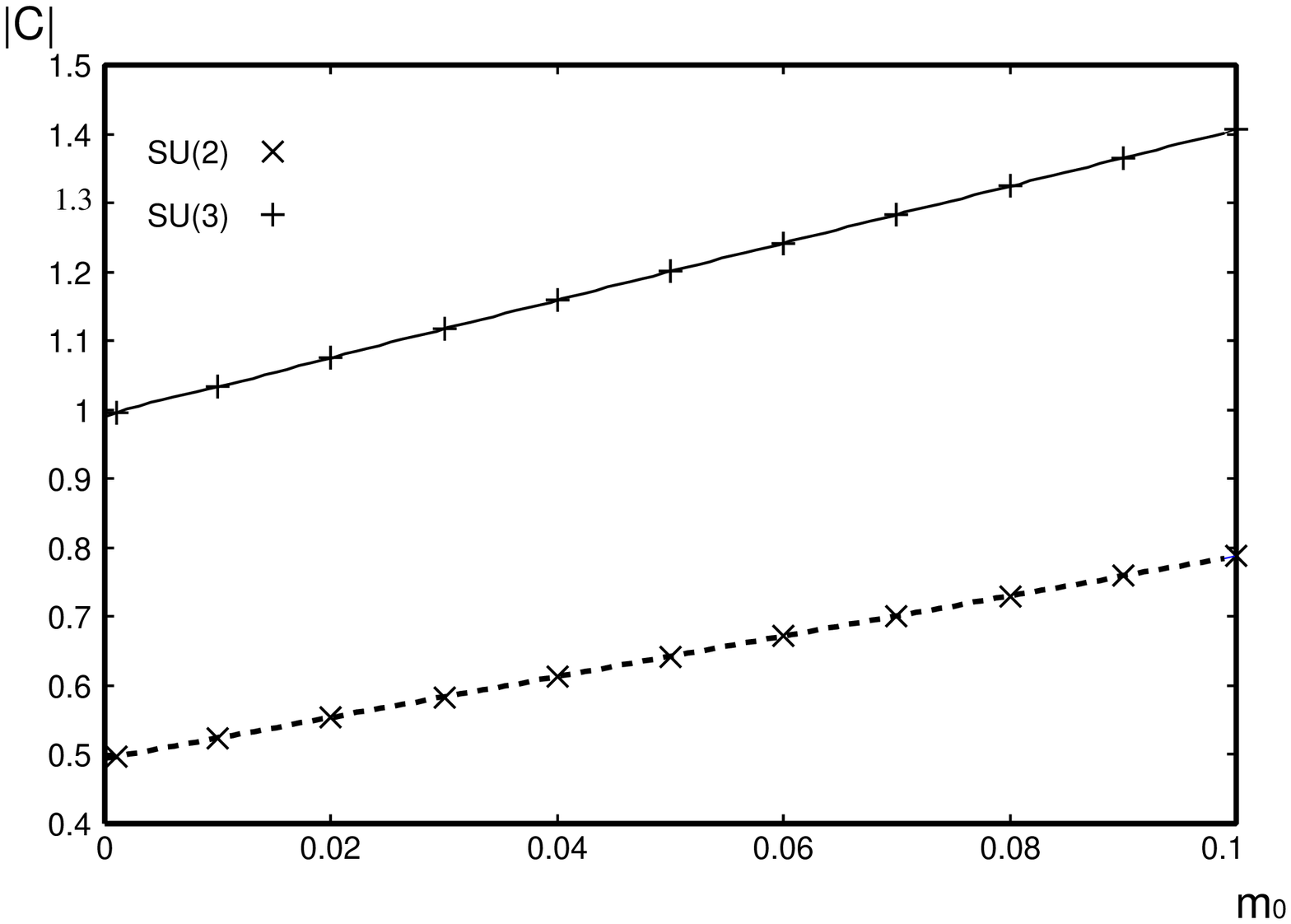}
\\
Fig.1

The absolute values of the condensate for $SU(2)$ and $SU(3)$ colors 
are plotted as the function 
of the fermion mass $m_0$ in the very small mass regions. 
The solid and dashed lines are shown to guide the eyes. 

\newpage
\includegraphics[width=.75\linewidth]{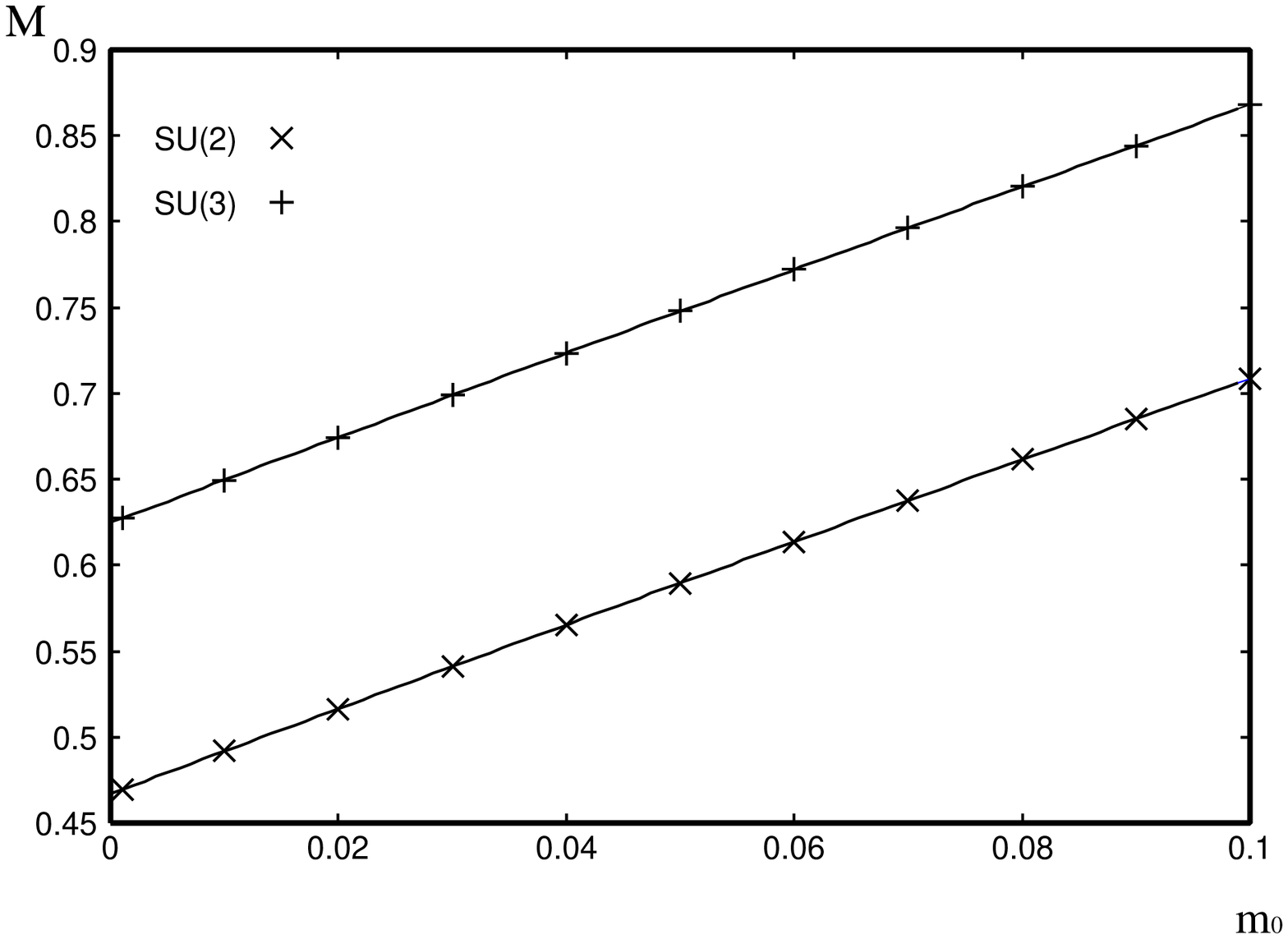}
\\
Fig.2

The boson masses for $SU(2)$ and $SU(3)$ colors are plotted as the function 
of the fermion mass $m_0$ in the very small mass regions. The solid lines 
are shown to guide the eyes.

\newpage
\includegraphics[width=.75\linewidth]{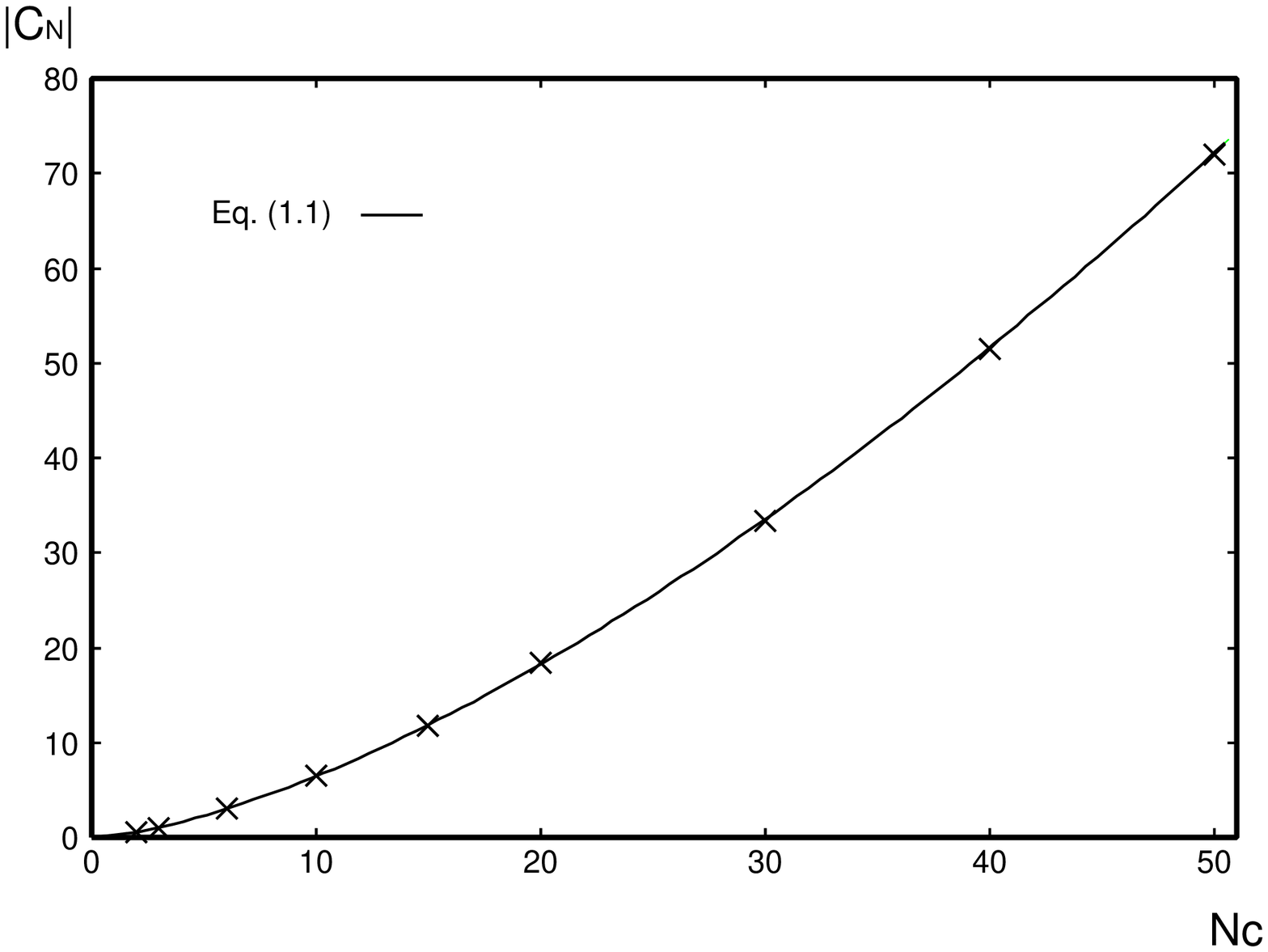}
\\
Fig.3

The absolute values of the condensate for $SU(N_c)$ colors 
are plotted as the function of $N_c$. The crosses are the calculated 
values while the solid line is the prediction of eq.(1.1). 

\newpage
\includegraphics[width=.75\linewidth]{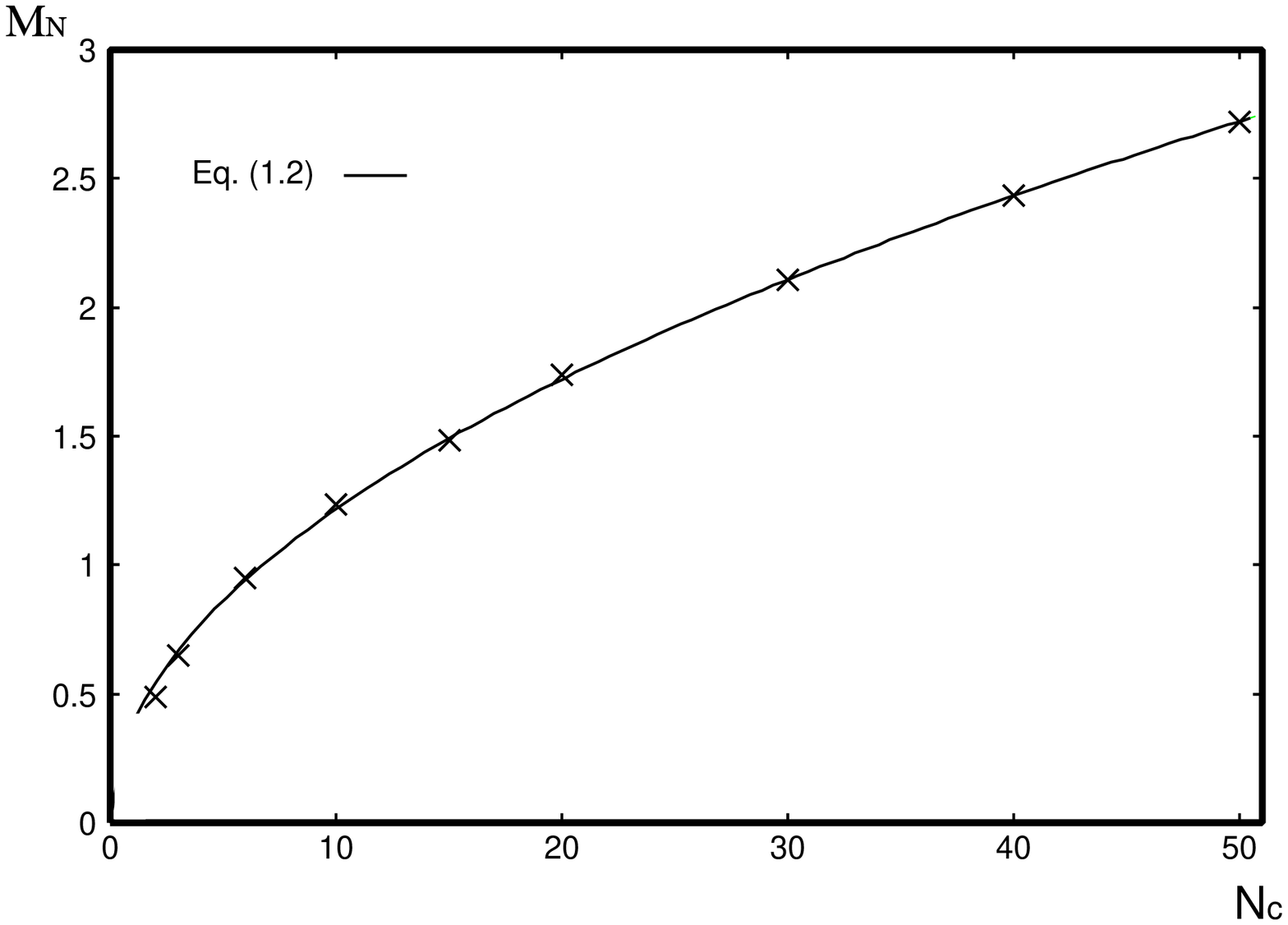}
\\
Fig.4

The boson masses for $SU(N_c)$ colors with the massless fermion 
are plotted as the function of $N_c$. The crosses are the calculated 
values while the solid line is the prediction of eq.(1.2). 

\newpage
\includegraphics[width=.75\linewidth]{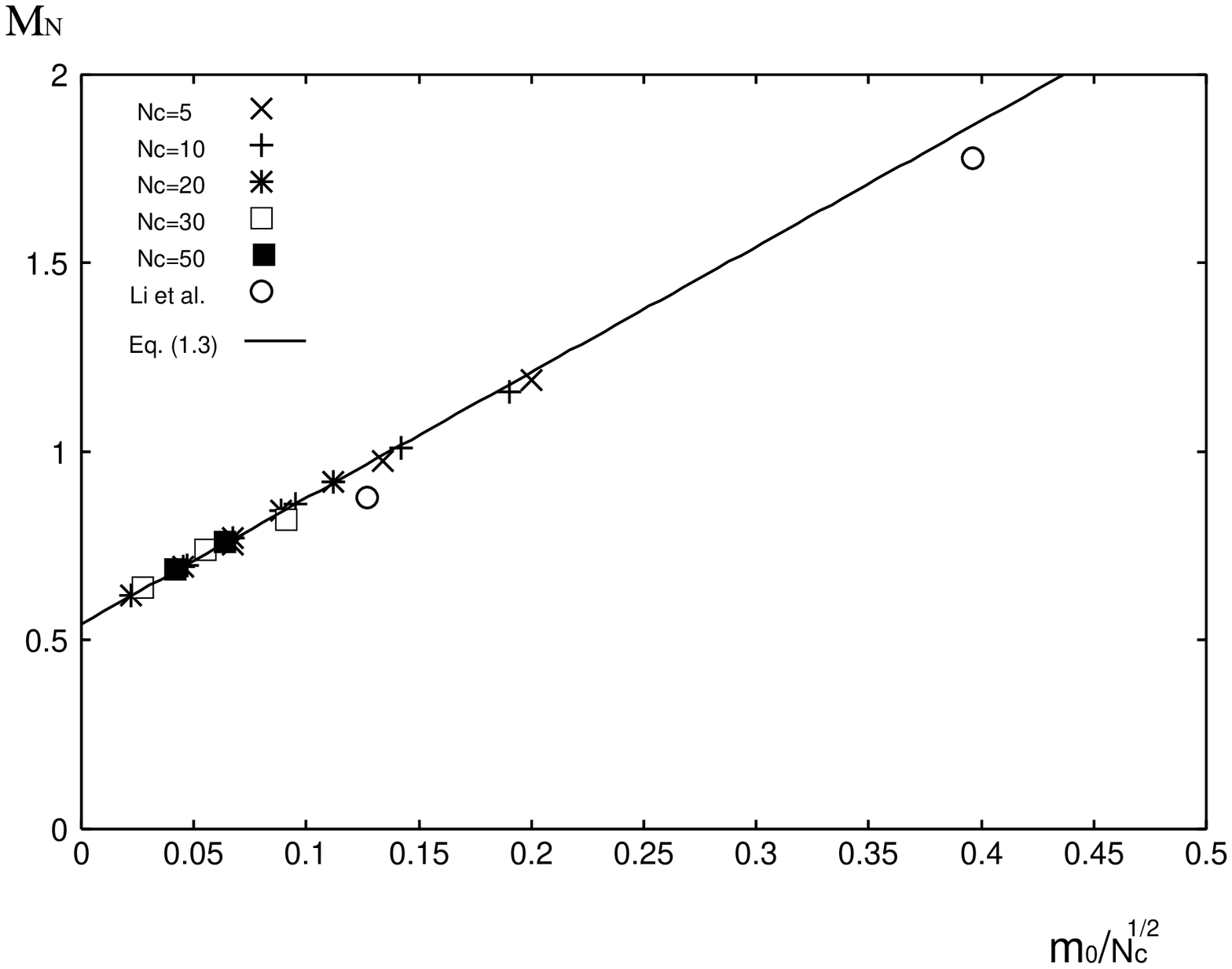}
\\
Fig.5

The boson masses in units of $\sqrt{{N_cg^2\over{2\pi}}}$ 
for $SU(N_c)$ colors with the massive fermion 
are plotted 
as the function of $m_0/\sqrt{N_c}$. The crosses, 
circles and squares are the calculated 
values while the solid line is the prediction of eq.(1.3).


\newpage



\begin{thebibliography}{99}

\bibitem{q01} 
G. 't Hooft, Nucl. Phys. {\bf B75} (1974), 461

\bibitem{q022} 
K. Hornbostel, S.J. Brodsky, and H.C. Pauli,  Phys. Rev. {\bf D41} (1990), 3814

\bibitem{q02} 
S.J. Brodsky, H.C. Pauli, and S.S. Pinsky, Phys. Rep. {\bf 301} (1998), 299 

\bibitem{q21} 
M. Li, Phys. Rev. {\bf D34} (1986), 3888

\bibitem{q210} 
M. Li, L. Wilets, and M.C. Birse, J. Phys. {\bf G13} (1987), 915


\bibitem{q211} 
I. Bars and M.B. Green, Phys. Rev. {\bf D17} (1978), 537

\bibitem{q22}
E. Abdalla and N.A. Alves,  Ann. Phys. {\bf 277} (1999), 74

\bibitem{q23} 
L.L.Salcedo, S. Levit, and J.W. Negele, Nucl. Phys. {\bf B361} (1991), 585


\bibitem{q3}
N.D. Mermin and H. Wagner,  Phys. Rev. Lett. {\bf 17} (1966), 1133

\bibitem{q4}
S.Coleman, Comm. Math. Phys. {\bf 31} (1973), 259

\bibitem{q05}
A.B. Zhitnitsky,  Phys. Lett. {\bf B165} (1985), 405

\bibitem{q06}
M. Burkardt, Phys. Rev. {\bf D53} (1996), 933

\bibitem{q03} 
F. Lenz, M. Thies, and K. Yazaki, Phys. Rev. {\bf D63} (2001), 045018

\bibitem{q04}
M. Burkardt, F. Lenz, and M. Thies,  Phys. Rev. {\bf D65} (2002), 125002


\bibitem{q11}
 T. Tomachi and T. Fujita,  Ann. Phys. {\bf 223} (1993), 197

\bibitem{q1} 
J. Goldstone, Nuovo Cimento, {\bf 19} (1961), 154

\bibitem{q2}
 J. Goldstone, A. Salam and S. Weinberg, Phys. Rev. {\bf 127} (1962), 965


\bibitem{q5}
 M. Faber and A.N. Ivanov, Eur. Phys. J. {\bf C20} (2001), 723


\bibitem{q9}
M. Hiramoto and T. Fujita,  "No massless boson in chiral symmetry breaking 
in Thirring and NJL models", hep-th/0306083

\bibitem{q10}
 M. Hiramoto and T. Fujita,  Phys. Rev. {\bf D66} (2002), 045007

\bibitem{q12}
 T. Fujita and A. Ogura, Prog. Theor. Phys. {\bf 89} (1993), 23

\bibitem{q13}
 T. Fujita, Y. Sekiguchi and K. Yamamoto, Ann. Phys. {\bf 255} (1997), 204

\bibitem{q15}
T. Fujita, A. Ogura and T. Tomachi,  Ann. Phys. {\bf 237} (1995), 12

\end{thebibliography}
\end{document}